\documentclass[twocolumn,epjc3]{svjour3}
\smartqed  % flush right qed marks, e.g. at end of proof
\RequirePackage{graphicx}
\begin{document}

\title{Improved constraints on the coupling constants of
axion-like particles to nucleons
 from recent Casimir-less experiment}

\titlerunning{Constraints on the coupling constants of
axion-like particles to nucleons}

\author{
G.~L.~Klimchitskaya\thanksref{addr1,addr2}\and
V.~M.~Mostepanenko\thanksref{e1,addr1,addr2}}
\authorrunning{G.~L.~Klimchitskaya and V.~M.~Mostepanenko}
\thankstext{e1}{e-mail: vmostepa@gmail.com}
\institute{
Central Astronomical Observatory
at Pulkovo of the Russian Academy of Sciences,
St.Petersburg, 196140, Russia \label{addr1}
\and
Institute of Physics, Nanotechnology and
Telecommunications, St.Petersburg State
Polytechnical University, St.Petersburg, 195251, Russia
\label{addr2}
}
\date{Received: date / Revised version: date}
\maketitle

\abstract{
We obtain improved constraints on the coupling constants
of axion-like particles to nucleons from a recently performed
Casimir-less experiment. For this purpose, the differential
force between a Au-coated sphe\-re and either  Au or Si sectors
of a rotating disc, arising due to two-axion exchange, is
calculated. Over a wide region of axion masses from
$1.7\times 10^{-3}\,$eV to 0.9\,eV the obtained constraints
are stronger up to a factor of 60 than the previously known
ones following from the Cavendish-type experiment and
measurements of the effective Casimir pressure.
%
%\PACS{
%     {14.80.Va}{discribing text of that key}   \and
%      {12.20.Fv}{discribing text of that key} \and
%      {14.80.-j}{discribing text of that key}
%     } % end of PACS codes
} %end of abstract
\section{Introduction}

It is common knowledge that the proper QCD axions are pseudoscalar particles which
appear as a consequence of breaking the Peccei-Quinn symmetry \cite{1} proposed
to resolve the problem of strong {\it CP} violation in QCD. After the prediction
of axions \cite{2,3}, a lot of experimental and theoretical work has been done
on their search and investigation of their role in elementary particle physics,
astrophysics and cosmology \cite{4,5,6,7,8,9,10,11,11a}.
At the moment, the originally introduced QCD axions, which are
pseudo-Nambu-Goldstone bosons, are constrained to a very narrow band in parameter
space \cite{11}, and many types of so-called {\it axion-like particles} are
proposed in different models (see, for instance, the {\it hadronic axions} \cite{12,13}
and the {\it GUT axions} \cite{14,15}).

Axion-like particles interact with photons, electrons and nucleons. Many searches of
these particles are based on the use of helioscopes and haloscopes \cite{16}.
The helioscopes are created for registration of axion-like particles generated in
the sun \cite{17,18,18a,19}.
The haloscopes exploit an idea that axion-like particles are possible constituents
of dark matter \cite{4,20} and fill all the space around us. Then, their coupling
to photons can be detected using a cryogenic microwave cavity in strong magnetic
field \cite{11,21,22}. Many constraints on the parameters of axion-like particles
were obtained also from different astrophysical processes (see, for instance,
\cite{23,24,25,26,27,28,29,30}).
Thus, from the neutrino data of supernova SN~1987A the coupling constant of the 
hadronic axions to nucleons was shown to be less than $10^{-10}$ or larger than 
$10^{-3}$ with a narrow allowed region in the vicinity of $10^{-6}$ \cite{24}. 
Stellar cooling by the emission of hadronic axions leads to a conclusion that 
for hadronic axions this interaction constant is less than $3\times 10^{-10}$ 
\cite{25,30}. It is noted \cite{25}, however, that the emission rate suffers from 
significant uncertainties related to dense nuclear matter effects. New strong 
limit on an axion mass and, thus, on an axion-to-nucleon interaction constant 
(which are connected for hadronic axions) was obtained from direct Chandra 
observations of the surface temperature of isolated neutron star in Cassiopeia~A 
and its cooling scenario \cite{32a}.

The model-independent laboratory constraints on the coupling constants of
axion-like particles with nucleons were obtained from neutron physics \cite{31,32},
E{\"{o}}t\-vos- and Cavendish-type experiments \cite{33,34,35}, and from measurements
of the Casimir and Casimir-Polder force \cite{36,37,38,39}.
These constraints cover a wide range of masses of axion-like particles from
$10^{-10}\,$eV to 20\,eV. As was shown in \cite{36,37,38,39} (see also \cite{40} for a
review), measurements of the Casimir interaction lead to stronger constraints
on the coupling constants of axions to nucleons than those obtained from the Cavendish-type
experiments. This corresponds to separation distances
between the test bodies where the Casimir interaction becomes stronger than the
gravitational one.

In this paper, we obtain improved constraints on the coupling constants of 
axion-like particles  
to a proton and a neutron following from the recently performed Casimir-less
experiment \cite{41}. This is the differential force measurement between a
Au-coated sphere and either a Au sector  or a Si sector of the structured disc
deposited on a Si substrate and covered by the overlayers of Cr and Au.
In such a manner, the contribution of the Casimir force to the differential
signal is subtracted, and the measurement result is determined solely by  a
difference in the forces due to exchange of some hypothetical particles.
By achieving the unprecedented force sensitivity of approximately $10^{-16}\,$N,
an improvement of constraints on the Yukawa-type corrections to Newtonian
gravitation by a factor of $10^3$ was achieved \cite{41}.
The corrections of Yukawa type arise due to exchange of one scalar boson between
two atoms of the laboratory test bodies \cite{42} or from compact extra dimensions
with low-energy compactification scale \cite{43}.
Here, we use the same experimental results to improve the previously known
laboratory constraints on the coupling constants of axion-like particles to
nucleons. Taking into account that the test bodies are unpolarized \cite{41}
and the axion-like particles are pseudo-scalar, the additional axionic interaction
arises due to two-axion exchange between nucleons of the test bodies.
Here, we strengthen the axion-to-nucleon coupling constants up to a factor of 60
within the wide region of axion masses from $1.7\times 10^{-3}$ to 0.9\,eV.
All equations are written in the system of units with $\hbar=c=1$.

\section{Differential force between a sphere and a structured disc due to
two-axion exchange}

In the experiment \cite{41}, a Au-coated sapphire sphere of $R=149.3\,\mu$m
radius interacts in vacuum with either a Au sector or a Si sector of the structured
rotating disc of thickness $D=2.1\,\mu$m, and the difference in these interaction
forces is an immediately measured quantity. The structured disc was deposited on
a Si substrate and covered by the overlayers of Cr and Au of thicknesses
$d_{\,\rm Cr}=10\,$nm and $d_{\rm Au}=150\,$nm, respectively.
Note that in \cite{41} the concentric alternating strips of Au and Si have been
used rather than a sectoral structure. This, however, does not influence our
calculation of the differential force. An important point is that the thick
overlayer of Au results in equal Casimir forces when the sphere bottom is above
a Au sector or a Si sector. Thus, the Casimir force does not contribute to the
measured differential force. The latter is determined by possible hypothetical
interactions, such as the Yukawa-type correction to Newton's gravitational
law \cite{41} or the two-axion exchange between nucleons of the sphere and the
structured disc under consideration here (if both these attractive interactions
exist in nature and contribute to the measured differential force, the constraints
imposed on each of them by the measurement data would be even stronger than those
obtained in \cite{41} and in this paper).

In this section, we consider the homogeneous Au sphere interacting due to
two-axion exchange between nucleons with the structured Au/Si disc.
We assume the pseudoscalar character of the axion-nucleon interaction, which is
applicable to wide classes of axion-like particles, specifically, to all
GUT axions \cite{44} with no connection between their mass and their interaction 
constant. 
Note that the account of scalar coupling of axions to
fermions \cite{44} or interaction of axions with electrons could only slightly
increase the magnitude of a differential axionic force and, thus, only slightly
strengthen the obtained constraints (see \cite{32,45} for the constraints
on scalar interaction of axions with nucleons).

We assume that the coordinate plane $(x,y)$ coincides with the upper plane of
the disc and the $z$ axis is perpendicular to it. The origin of the coordinate
system is chosen below the bottom point of the sphere nearest to the disc.
Without loss of accuracy one can neglect by the finite size effects and consider
the disc of infinitely large area \cite{38,46}. The separation distance between
the sphere and the disk is $a$, so that the sphere center is at $z=a+R$.
The effective potential due to two-axion exchange between two nucleons situated
at the points $\mbox{\boldmath$r$}_1$ of the sphere and $\mbox{\boldmath$r$}_2$
of the disc  is given by \cite{33,47,48}
\begin{equation}
V_{kl}(|\mbox{\boldmath$r$}_1-\mbox{\boldmath$r$}_2|)=
-\frac{g_{ak}^2g_{al}^2\,m_a}{32\pi^3m^2}\,
\frac{K_1(2m_a|\mbox{\boldmath$r$}_1-
\mbox{\boldmath$r$}_2|)}{(\mbox{\boldmath$r$}_1-\mbox{\boldmath$r$}_2)^2}
.
\label{eq1}
\end{equation}
\noindent
Here, $g_{ak}$ and $g_{al}$ are the coupling constants of
interaction between an axion-like particle of mass $m_a$ and a proton
($k,\,l=p$) or a neutron ($k,\,l=n$),  the mean mass of a nucleon is
 $m=(m_n+m_p)/2$, and $K_1(z)$ is the modified
Bessel function of the second kind. Equation (\ref{eq1}) is
applicable under the condition
$|\mbox{\boldmath$r$}_1-\mbox{\boldmath$r$}_2|\gg 1/m$ which
is satisfied with large safety margin because in the experiment
$a>200\,$nm \cite{41}.

The additional force due to two-axion exchange, acting between a homogeneous
sphere $(s)$ and a homogeneous disc $(d)$, was found in \cite{37} by the
summation of microscopic forces determined by the potential (\ref{eq1}):
\begin{eqnarray}
&&
F_{\rm add}(a)=
\frac{\pi m_a}{m^2m_H^2}C_dC_s
\int_{a}^{2R+a}\!\!\!dz_1[R^2-(z_1-R-a)^2]
\nonumber\\
&&~~~~~~~~~~~~~~~~
\times
I(m_a,D,z_1),
\label{eq2}
\end{eqnarray}
\noindent
where
\begin{eqnarray}
&&
I(m_a,D,z_1)\equiv
\frac{\partial}{\partial z_1}\int_{-D}^{0}\!dz_2
\int_{0}^{\infty}\!\!\!\rho d\rho
\label{eq3}\\
&&~~~~~~~~~~~~~~
\times
\frac{K_1(2m_a\sqrt{\rho^2+(z_1-z_2)^2})}{\rho^2+(z_1-z_2)^2}\,.
\nonumber
\end{eqnarray}
\noindent
Here, the coefficient $C_{d,s}$ for the disc and sphere materials is defined as
\begin{equation}
C_{d,s}=\rho_{d,s}\left(\frac{g_{ap}^2}{4\pi}\,
\frac{Z_{d,s}}{\mu_{d,s}}+\frac{g_{an}^2}{4\pi}\,
\frac{N_{d,s}}{\mu_{d,s}}\right),
\label{eq4}
\end{equation}
\noindent
where $\rho_{d,s}$ is the disc and sphere densities, and $Z_{d,s}$ and
$N_{d,s}$ are
the number of protons and the mean number of neutrons in the
atoms of a disc and a sphere. The quantities $\mu_{d,s}$
are defined as $\mu_{d,s}=m_{d,s}/m_H$ where
$m_{d,s}$  are the mean  masses of the disc and sphere atoms,
and $m_H$ is the mass of an atomic hydrogen, respectively.
The values of $Z/\mu$ and $N/\mu$ for many elements
with account of their isotopic
composition can be found in \cite{42}.

The quantity $I$ defined in (\ref{eq3}) can be equivalently represented in
the form \cite{37}
\begin{eqnarray}
&&
I(m_a,D,z_1)=-\int_{1}^{\infty}\!du\frac{\sqrt{u^2-1}}{u}
e^{-2m_auz_1}
\label{eq5}\\
&&~~~~~~~~~~~~~
\times
\left(1-e^{-2m_auD}\right).
\nonumber
\end{eqnarray}
\noindent
Using (\ref{eq2}) and (\ref{eq5}), the magnitude of the
differential additional force
arising from the alternate interaction of a sphere with Au and Si sectors of
the structured disc is equal to
\begin{eqnarray}
&&
|\Delta F_{\rm add}(a)|=
\frac{\pi m_a}{m^2m_H^2}C_s(C_{\rm Au}-C_{\rm Si})
\label{eq6}\\
&&~~~
\times
\int_{1}^{\infty}\!du\frac{\sqrt{u^2-1}}{u}\left(1-e^{-2m_auD}\right)
\nonumber\\
&&~~~~~~
\times
\int_{a}^{2R+a}\!\!dz_1e^{-2m_auz_1}[R^2-(z_1-R-a)^2].
\nonumber
\end{eqnarray}
\noindent
Here, $C_{\rm Au}$ and $C_{\rm Si}$ are defined by (\ref{eq4}) for Au and Si
disc materials, respectively.

It is convenient to introduce the new integration variable $t=z_1-a$ in the integral
with respect to $z_1$. After the integration with respect to $t$ is performed,
the differential force (\ref{eq6}) is given by
\begin{eqnarray}
&&
|\Delta F_{\rm add}(a)|=
\frac{\pi }{2m_am^2m_H^2}C_s(C_{\rm Au}-C_{\rm Si})
\label{eq7}\\
&&~~
\times\!\!
\int_{1}^{\infty}\!\!du\frac{\sqrt{u^2-1}}{u^3}e^{-2m_aua}\left(1-e^{-2m_auD}\right)
\Phi(R,m_au),
\nonumber
\end{eqnarray}
\noindent
where
\begin{equation}
\Phi(r,z)=r-\frac{1}{2z}+e^{-2rz}\left(r+
\frac{1}{2z}\right).
\label{eq8}
\end{equation}
\noindent
Note that equation (\ref{eq7}) is the exact one. The differential force due to two-axion
exchange can be obtained also using the proximity force approximation \cite{49}, where
the spherical surface is replaced with infinitesimally small plane plates parallel
to the disc. This again results in (\ref{eq7}), but with $\Phi=R$. Such an approximate
expression is only applicable under a condition $R\gg m_a^{-1}$.

In the next section, we apply the expression (\ref{eq7}) to calculate the differential
force in the configuration of experiment \cite{41} and obtain constraints on the
coupling constants of axion-like particles to nucleons.

\section{Improved constraints on the
coupling constants of axions to nucleons}

Now we take into account that in the experiment \cite{41} the sphere was not homogeneous.
It was made of sapphire (Al${}_2$O${}_3$) and covered with the layers of Cr and Au of
thicknesses $\Delta_{\rm Cr}=10\,$nm and $\Delta_{\rm Au}=250\,$nm, respectively.
Thus, the differential force between a sapphire core and a structured disc can be
calculated by (\ref{eq7}) with $C_s=C_{\rm Al_2O_3}$ where $R$ is replaced with
$R-\Delta_{\rm Au}-\Delta_{\rm Cr}$. Then one should add to the obtained result the
differential forces between the structured plate and each of two spherical envelopes
of external radia $R-\Delta_{\rm Au}$ and $R$ of thicknesses $\Delta_{\rm Cr}$
and $\Delta_{\rm Au}$, made of Cr and Au, respectively.
These differential forces are calculated similarly using (\ref{eq7}). For instance,
the differential force due to a Au envelope is found by subtracting from (\ref{eq7})
with $C_s=C_{\rm Au}$ the differential force due to a Au sphere of radius
$R-\Delta_{\rm Au}$ placed at a separation $a+\Delta_{\rm Au}$ from the structured
plate:
\begin{eqnarray}
&&
|\Delta F_{\rm add}^{\rm Au}(a)|=
\frac{\pi }{2m_am^2m_H^2}C_{\rm Au}(C_{\rm Au}-C_{\rm Si})
\label{eq9}\\
&&~~
\times\!\!
\int_{1}^{\infty}\!\!du\frac{\sqrt{u^2-1}}{u^3}e^{-2m_aua}\left(1-e^{-2m_auD}\right)
\nonumber \\
&&~~~~~~~
\left[\Phi(R,m_au)-e^{-2m_au\Delta_{\rm Au}}
\Phi(R-\Delta_{\rm Au},m_au)\right].
\nonumber
\end{eqnarray}

By adding up the differential forces from the sapphire core and two spherical layers,
we arrive at the following result valid in the experimental configuration \cite{41}:
\begin{eqnarray}
&&
|\Delta F_{\rm add}^{\rm exp}(a)|=
\frac{\pi }{2m_am^2m_H^2}(C_{\rm Au}-C_{\rm Si})
\label{eq10}\\
&&~~
\times\!\!
\int_{1}^{\infty}\!\!du\frac{\sqrt{u^2-1}}{u^3}e^{-2m_aua}\left(1-e^{-2m_auD}\right)
X(m_au),
\nonumber
\end{eqnarray}
\noindent
where
\begin{eqnarray}
&&
X(z)\equiv C_{\rm Au}
\left[\Phi(R,z)-e^{-2z\Delta_{\rm Au}}
\Phi(R-\Delta_{\rm Au},z)\right]
\label{eq11} \\
&&
+C_{\rm Cr}e^{-2z\Delta_{\rm Au}}
\left[\vphantom{e^{-2z\Delta_{\rm Cr}}}
\Phi(R-\Delta_{\rm Au},z)\right.
\nonumber \\
&&~~~~~~~~\left.
-e^{-2z\Delta_{\rm Cr}}
\Phi(R-\Delta_{\rm Au}-\Delta_{\rm Cr},z)\right]
\nonumber \\
&&
+C_{\rm Al_2O_3}e^{-2z(\Delta_{\rm Au}+\Delta_{\rm Cr})}
\Phi(R-\Delta_{\rm Au}-\Delta_{\rm Cr},z).
\nonumber
\end{eqnarray}
\noindent
Note that the quantities $C$ for Au, Cr, and Al${}_2$O${}_3$ are
defined by (\ref{eq4}). The values of $Z/\mu$ and $N/\mu$  for atoms
of Au, Cr and Si, and for a molecule of Al${}_2$O${}_3$, as well as
$\rho$ for these materials, are presented in Table~I of \cite{38}.

Now we are in a position to obtain constraints on the parameters of
axion-like particles following from the measurement results of experiment
\cite{41}. Recall that neither the Si substrate under the structured disc nor
the Au and Cr overlayers contribute to the measured differential force
which is given by (\ref{eq10}) and (\ref{eq11}). However, when using the
results of \cite{41}, it should be remembered that the experimental
separation distances between the rotating disc and the sphere are equal to
$z=a-d_{\rm Au}-d_{\,\rm Cr}=a-160\,$nm.

In the experiment \cite{41} no differential force was observed. This means that
the quantity (\ref{eq10}) arising due to two-axion exchange between the test
bodies was less than the minimum detectable force
\begin{equation}
\Delta F_{\rm add}^{\rm \,exp}(a)\leq\delta F(a).
\label{eq12}
\end{equation}
\noindent
According to Fig.~3 of \cite{41}, at separation distances $z=200$, 400, 700, and
1000\,nm (this corresponds to $a=360$, 560, 860, and 1160\,nm) the minimum detectable
force was equal to $\delta F=0.2$, 0.09, 0.12, and 0.12\,fN, respectively.
These values were determined at the 95\% confidence level. We have found numerically
the values of the axion-to-nucleon coupling constants $g_{ap}$, $g_{an}$ and masses
$m_a$ satisfying the inequality (\ref{eq12}) with $\Delta F_{\rm add}^{\rm \,exp}$
given by (\ref{eq10}) and (\ref{eq11}). The most strong constraints were obtained
at $a=560\,$nm.

%%%%%%%__Figure_1___%%%%%%%%
\begin{figure}[t]
\vspace*{-5.6cm}
\resizebox{0.6\textwidth}{!}{%
\hspace*{-2.7cm} \includegraphics{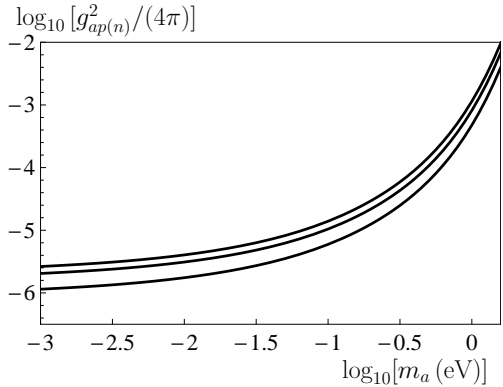}
}
\vspace*{-6cm}
\caption{Constraints on the coupling constants of axion-like
particles to
a proton or a neutron obtained from the
Casimir-less experiment \cite{41} are shown as
functions of the axion mass.
The lines from bottom to top are plotted
under the conditions
$g_{ap}^2=g_{an}^2$, $g_{an}^2\gg g_{ap}^2$, and
$g_{ap}^2\gg g_{an}^2$, respectively.
The regions of the plane above each line are excluded and
below each line are allowed.}
\label{fig:1}       % Give a unique label
\end{figure}
%%%%%%%%%%%%%%%%%%%%%%%%
The computational results for allowed and excluded values of the coupling constants
$g_{ap(n)}^2/(4\pi)$ as functions of the axion mass $m_a$ are presented in
Fig.~1. The three lines from bottom to top are plotted
under the conditions
$g_{ap}^2=g_{an}^2$, $g_{an}^2\gg g_{ap}^2$, and
$g_{ap}^2\gg g_{an}^2$, respectively.
The regions of the plane $(m_a,g_{ap(n)}^2)$ above each line are excluded
by the experimental results, and the regions below each line are allowed.
As can be seen in Fig.~1, with increasing $m_a$ the strength of the obtained
constraints quickly decreases, and they become not competitive.

%%%%%%%__Figure_2___%%%%%%%%
\begin{figure}[t]
\vspace*{-3.6cm}
\resizebox{0.6\textwidth}{!}{%
\hspace*{-3.cm} \includegraphics{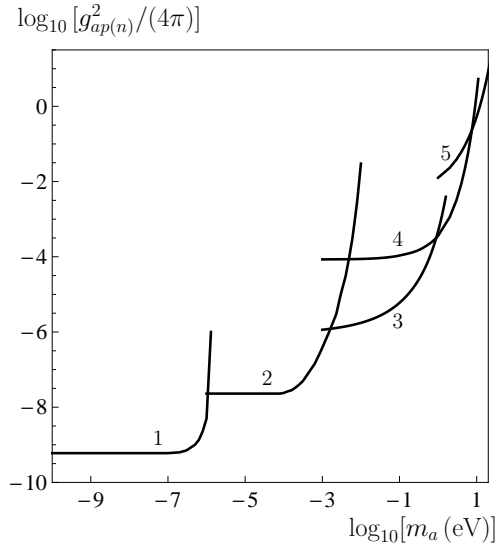}
}
\vspace*{-6cm}
\caption{Constraints on the coupling constants of
axion-like particles to
a proton and  a neutron obtained
under the condition $g_{ap}^2=g_{an}^2$
from the  magnetometer measurements \cite{31} (line 1),
from the Cavendish-type experiment \cite{34,35}
(line 2),  in this work from the Casimir-less experiment
\cite{41} (line 3), from measurements of the effective
Casimir pressure \cite{38,50,51} (line 4), and from
measurements of the lateral Casimir force between corrugated
surfaces \cite{39,52,53} (line 5).
The regions of the plane above each line are excluded and
below each line are allowed.
}
\label{fig:2}       % Give a unique label
\end{figure}
%%%%%%%%%%%%%%%%%%%%
It is interesting to compare the constraints of Fig.~1 with previously obtained
strongest laboratory constra\-ints on axion-like particles. This comparison is
 presented in Fig.~2 under the most reasonable condition $g_{ap}=g_{an}$
\cite{33}. In Fig.~2, the line 1 shows the constraints found from the
magnetometer measurements using spin-po\-la\-rized K and ${}^3$He atoms \cite{31}
(note also recent limits on a product of the pseudoscalar and scalar axion-to-nucleon 
interaction constants obtained \cite{51a,51b} from the magnetometer experiment with 
${}^3$He and ${}^{129}$Xe atoms).
The line 2 shows the constraints derived \cite{35} from the Cavendish-type
experiment \cite{34}. The line 3 demonstrates the constraints obtained in this
paper from the Casimir-less experiment \cite{41}. This line reproduces the
lowest line of Fig.~1. The constraints found \cite{38} from measurements of the
effective Casimir pressure by means of a micromachined oscillator \cite{50,51} are
presented by the line 4. Finally, the line 5 shows the constraints derived \cite{39}
from measurements of the lateral Casimir force between sinusoidally corrugated
surfaces \cite{52,53}.
The regions of $(m_s,g_{ap(n)}^2)$  plane above all lines are excluded
by the experimental results, and the regions below each line are allowed.

As can be seen in Fig.~2, the obtained in this paper constraints of line 3
are significantly improved in comparison with 
the previously known ones in the wide region
of axion masses from $1.7\times 10^{-3}\,$eV to 0.9\,eV. In the region
from $1.7\times 10^{-3}\,$eV to $4.9\times 10^{-3}\,$eV our constraints
strengthen the constraints obtained \cite{35} from the Cavendish-type
experiment \cite{34} whereas in the region from $4.9\times 10^{-3}\,$eV to
0.9\,eV they are stronger than the constraints found \cite{38} from
measurements of the effective Casimir pressure \cite{50,51}.
The largest strengthening of previously known constraints by a factor of 60
holds for $m_a=4.9\times 10^{-3}\,$eV.  Thus, the Casimir-less experiment
leads to stronger constraints not only on the Yukawa-type corrections
to Newton's gravitational law \cite{41}, but on the coupling constants
of axion-like particles to nucleons as well.

\section{Conclusions and discussion}

In the foregoing we have used the experimental results of recent Casimir-less
experiment to derive the constraints on the axion-to-nucleon coupling constants.
The obtained constraints strengthen the previously known ones following from the
Cavendish-type experiment and from measurements of the effective Casimir pressure.
The strengthening up to a factor of 60 is achieved
over the wide region
of axion masses from $1.7\times 10^{-3}\,$eV to 0.9\,eV.

It should be remarked that the results of the Casimir-less experiment used here to
obtain the constraints on axion-nucleon interaction are unambiguous in the sense that
they do not depend on any theory of the Casimir force. This is different, for instance,
from the constraints obtained \cite{38} from the measure of
agreement between measured
effective Casimir pressure and theory (see the line 4 in Fig.~2).
It is well known that precise experiments on measuring the Casimir interaction
between metallic surfaces agree with the extrapolation of a dielectric function to
zero frequency using the plasma model, whereas a literally understood theory suggests
to use the Drude model at low frequencies \cite{49,50,51,55,56,57,58}.
Recently the decisive experiment has been proposed \cite{59}, where the theoretical
predictions of both approaches differ by a factor of order $10^3$. The first
measurements performed in the framework of this proposal are in favor of the plasma model
extrapolation \cite{60}. However, a fundamental understanding of physical mechanisms
behind this problem is still missing. Because of this, the obtained here constraints,
which are independent on theory-experiment comparison for the Casimir forces, are of
particular value.

In the past, both measurements of the Casimir interaction and the Casimir-less
experiment were used for obtaining constraints on the Yukawa-type corrections to
Newtonian gravity \cite{41,49,50,51,61,62,63,64,65,66}.
In references \cite{36,37,38,39,40} and in this paper it is shown that the same
experiments also lead to competitive constraints on the coupling constants of
an axion-to-nucleon interaction. It should be remembered, however, that the
Yukawa potential arises due to exchange of one scalar particle, whereas the
spin-independent potential (\ref{eq1}), used to constrain the parameters of
axion-like particles, results from the exchange of two particles. 
The latter makes the obtained
constraints relatively weaker. Thus, in the future it seems promising to perform
measurements of the Casimir interaction and Casimir-less experiment using the
polarized (magnetized) test bodies. In this way one could obtain much stronger
constraints on the parameters of axion-like particles by exploiting spin-dependent
interaction potential arising due to one-particle exchange between nucleons.

%%%%%%%%%%%%%%%%%%%%%%%%%%%%%%%%%%
%%%%%%%%%
\begin{acknowledgement}
%{\it Acknowledgments.}
The authors are grateful to R.\ S.\ Decca for
useful information about his experiment.
\end{acknowledgement}
%%%%%%%%%%%%%%%%%%%%%%%%%%%%

%%%%%%%%%
\end{document}